\documentstyle[aps,prl,twocolumn,epsf,float]{revtex}
\begin{document}

\twocolumn[\hsize\textwidth\columnwidth\hsize\csname @twocolumnfalse\endcsname

\title{
Inhomogeneous States of Nonequilibrium Superconductors: \\
Quasiparticle Bags and Antiphase Domain Walls
}

\author{
M.I.~Salkola and J.R.~Schrieffer
}
\address{
NHMFL and Department of Physics, Florida State University, 
Tallahassee, Florida 32310}

\date{December 19, 1997}
\maketitle

\begin{abstract}

Nonequilibrium properties of short-coherence-length $s$-wave superconductors 
are analyzed in the presence of extrinsic and intrinsic inhomogeneities. 
In general, the lowest-energy configurations of quasiparticle excitations 
are topological textures where quasiparticles segregate into antiphase 
domain walls between superconducting regions whose order-parameter
phases differ by $\pi$. Antiphase domain walls can be probed by various 
experimental techniques, for example, by optical absorption and NMR. 
At zero temperature, quasiparticles seldom appear as self-trapped bag states. 
However, for low concentrations of quasiparticles, they may be stabilized 
in superconductors by extrinsic defects.

\

\noindent PACS numbers: 74.80.-g, 74.20.-z, 74.25.Jb, 74.25.G7

\end{abstract}

\

\

]

\section{Introduction}

In general, studies of superconductors emphasize their equilibrium properties 
as probed by linear response. Equally important are conditions where 
the superconductor is driven far from equilibrium. 
A nonequilibrium state may be achieved, for example, by photoexciting 
quasiparticles \cite{testardi,sai,hu} or injecting them into the superconductor 
through a tunnel junction \cite{iguchi,fuchs,dynes}. These experiments 
have revealed a variety of interesting phenomena, ranging from first-order 
superconductor-metal transitions, to various instabilities and spatially 
inhomogeneous states with a laminar structure where either superconducting 
phases with distinct energy gaps or superconducting and normal phases coexist. 

At finite temperature, quasiparticle dynamics in the nonequilibrium
state is dominated by scattering with phonons and decay with phonon 
emission. These processes are characterized by the scattering time 
$\tau_s$ and the lifetime $\tau_\ast$. While these time scales are 
long enough compared to $\hbar/\Delta_0$ so that the superconducting 
energy gap $\Delta_0$ is sharply defined\cite{bob}, they are also 
typically of the same order of magnitude, $\tau_{s} \sim \tau_{\ast}$ \cite{doug}. 
As a consequence, it is not obvious that the quasiparticles will 
equilibrate to a meta-stable state even under steady-state conditions. 
In contrast, when the lifetime of quasiparticles is the longest time scale, 
quasiparticles will reach such a state making it possible for new phenomena 
to emerge. A meta-stable state is obtained, if the system has a symmetry 
group which makes it possible for the excited and ground states to 
transform according to different irreducible representations so that 
the quasiparticle excitations cannot decay. It is clear that in 
the absence of spin-orbit coupling the only practically 
meaningful symmetry group is spin-rotational symmetry, because in 
the superconducting phase broken gauge symmetry destroys the charge 
conservation. Quasiparticle excitations whose spins are aligned along 
the same direction may be obtained by using a ferromagnetic metal as 
a source \cite{SK}.

\begin{floating}[t]
\narrowtext
\begin{figure}
\setlength{\unitlength}{1truecm}
\epsfxsize=7.5cm
\ \ \ \epsffile{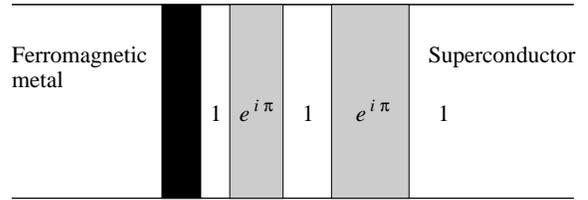}
\vskip 0.5truecm
\caption[]{\small
Schematic description of a tunnel-junction experiment where spin-polarized
quasiparticles are injected from a ferromagnetic metal through an
insulating barrier to a superconductor. In the superconductor, injected 
quasiparticles have formed a laminar structure of intervening superconducting 
domains separated by antiphase domain walls between regions where the neighboring 
order-parameter phases are shifted by $\pi$.
}
\end{figure}
\vskip -.5truecm\end{floating}

In this Note, we examine various meta-stable configurations of quasiparticles 
and their signatures that might develop when spin-polarized 
quasiparticles are excited in $s$-wave superconductors. 
Our most important finding is that superconductors are unstable against 
a formation of antiphase domain walls into which the quasiparticles 
localize and that the local structure and nonuniform spin density
makes these topological textures accessible to various experimental probes. 
In particular, they produce a distinctive optical absorption spectrum 
that may serve as a unique signature of their presence. In addition, 
any probe that is sensitive to a local magnetization would lend further 
support. For example, NMR and $\mu$SR might be suitable for this purpose.
Figure 1 illustrates schematically a tunnel-junction experiment
for generating and detecting antiphase domain-wall textures. 
Similar experimental construction has been suggested to demonstrate that
spin and charge are transported by separate quasiparticle excitations
in a superconductor \cite{KR}.
We also consider quasiparticle bags which are non-topological
states of quasiparticles associated with a local suppression of the order
parameter that may appear in the presence of defects when the quasiparticle 
density is small enough. 

Our work is partially motivated by the fact that only few studies exist
on self-trapped quasiparticle states in superconductors \cite{as,coffey} 
and that either quasiparticle-bag or antiphase domain-wall excitations 
are usually considered as a curiosity and often disregarded as unphysical 
\cite{coffey,schulz}. Our purpose is to address the question of their existence 
in mean-field approximation and to examine possible experimental
implications by focusing on quasi-one and two-dimensional superconductors 
which are realized in wires and films. Analogous questions have been studied 
in the context of antiferromagnets, although no detailed predictions regarding 
superconductors have been made \cite{schulz,verges}.

\section{Formalism}

Our starting point in describing quasiparticle excitations in an $s$-wave 
superconductor is the lattice formulation of electrons hopping between
nearest-neighbor sites and interacting via an effective two-particle 
interaction,
\begin{eqnarray}
H& = & -\mbox{$1\over 4$}W \sum_{\langle {\bf Rr}\rangle \sigma}
\psi_\sigma^\dagger({\bf R}+{\bf r}) \psi_\sigma({\bf R})
-\mu \sum_{\bf R \sigma} n_\sigma({\bf R}) \nonumber \\
& & \mbox { }+ U\sum_{{\bf R}} n_\uparrow({\bf R}) n_\downarrow({\bf R}).
\end{eqnarray}
Here, $\psi_\sigma({\bf r})$ is the electron operator with spin $\sigma$,
$\langle {\bf Rr}\rangle$ denotes nearest-neighbor sites separated by $\bf r$,
$W$ is the half bandwidth (on a square lattice), $\mu$ is the chemical potential,
and the operator $n_\sigma({\bf r})=\psi^\dagger_\sigma({\bf r}) 
\psi_\sigma({\bf r})$ is the conduction electron number density for spin 
$\sigma$.
The strength of the pairing interaction $U$ ($<0$) is
assumed to be intermediate so that the mean-field approximation
gives a qualitatively reliable description of the superconducting ground state 
and the low-energy excitations. Specifically, consider a two-dimensional lattice 
model where electrons can interact with randomly distributed defects. 
These defects can either have a magnetic moment or be nonmagnetic. 
The model is defined by the effective Hamiltonian $H_{\rm eff}=H_{\rm 0} 
+ H_{\rm imp}$, where $H_{\rm 0}$ describes a BCS superconductor \cite{bcs} 
and $H_{\rm imp}$ is the contribution due to impurities. In the mean-field 
approximation,
\begin{eqnarray}
H_{\rm 0}& = & -\mbox{$1\over 4$}W \sum_{\langle {\bf Rr}\rangle}
\Psi^\dagger({\bf R}+{\bf r})\hat{\tau}_3 \Psi({\bf R}) - \mu \sum_{\bf R}
\Psi^\dagger({\bf R})\hat{\tau}_3 \Psi({\bf R}) \nonumber \\
& &\ \ \ \  \mbox{\ } - \sum_{\bf Rr} \Delta({\bf R})
\Psi^\dagger({\bf R})\hat{\tau}_1 \Psi({\bf R}),
\label{eq:bcs}
\end{eqnarray}
where $\Delta({\bf R})$ is the superconducting gap function and assuming 
that the pairing of electrons occurs in the spin-singlet channel. 
The operator $\Psi({\bf r}) = [\psi_{\uparrow}({\bf r}) \ 
\psi_{\downarrow}^\dagger({\bf r})]^T$ is a two-component Gor'kov-Nambu 
spinor, $\hat{\tau}_\alpha$ ($\alpha=1,2,3$) are 
the Pauli matrices for particle-hole degrees of freedom, and $\hat{\tau}_0$ is 
the unit matrix. 
In a translationally invariant system, the BCS Hamiltonian reduces to
\begin{equation}
H_{\rm 0} = \sum_{\bf k} \Psi^\dagger_{\bf k} (\epsilon_{\bf k}\hat{\tau}_3 - 
\Delta_0\hat{\tau}_1)\Psi_{{\bf k}},
\label{eq:clean}
\end{equation}
where $\Delta_{0}= \Delta({\bf R})$ 
and $\Psi_{\bf k} = (\psi_{{\bf k}\uparrow} \ 
\psi_{-{\bf k}\downarrow}^\dagger)^T$. The fermion operators in real and momentum 
spaces are related by the unitary transformation, $\psi_\sigma({\bf r})= N^{-1/2} 
\sum_{\bf k} \psi_{{\bf k}\sigma} e^{i{\bf k}\cdot {\bf r}}$, where $N$ is 
the number of sites in the system. For a square lattice with the nearest-neighbor 
hopping, the single-particle energy relative to the chemical potential in 
the normal state is $\epsilon_{\bf k} = -{1\over 2}W(\cos k_xa + \cos k_ya) - \mu$; 
$a$ is the lattice spacing. In a uniform $s$-wave superconductor, 
the energy spectrum of bare quasiparticle excitations is 
$E_{\bf k}= \sqrt{\epsilon^2_{\bf k} + \Delta_0^2}$. 
Allowing the excited quasiparticles to relax, the energy
spectrum and the order parameter must be modified, as will be discussed below.

The interaction between the conduction electrons and the impurities in 
the superconductor is given by the Hamiltonian 
\begin{equation}
H_{\rm imp}= \sum_{\bf r}[V({\bf r})n({\bf r}) + J{\bf S}({\bf r})\cdot 
{\bf s}({\bf r})],   \label{eq:imp}
\end{equation}
where $n({\bf r})=\sum_\sigma n_\sigma({\bf r})$ and 
${\bf s}({\bf r})= {1\over 2}\sum_{\sigma\nu} \psi^\dagger_\sigma({\bf r})
\hat{\bf \tau}_{\sigma\nu} \psi_\nu({\bf r})$ are the conduction 
electron number density and spin density operators.
In the case of point like impurities located at sites ${\bf r}_n$,
the potential (scalar) and  magnetic scattering terms have the forms
$V({\bf r})= \sum_n V_n\delta_{{\bf r}{\bf r}_n}$ and ${\bf S}({\bf r})= \sum_n{\bf S}_n
\delta_{{\bf r}{\bf r}_n}$. Typically, the distribution of impurities
is assumed to be random whereas the magnitude of scalar and magnetic 
scattering are constant, $V_n= V$ and $w= JS/2$, where 
$S= |{\bf S}_n|$. For later emphasis, it is useful to introduce here 
a particle-hole transformation generated by the operator $\hat{\tau}_1$:
\begin{equation}
\Psi({\bf r}) \rightarrow \Psi'({\bf r})= (-1)^{\bf r}\hat{\tau}_1\Psi({\bf r}). 
\label{eq:trans}
\end{equation}
At half filling ($\mu=0$), the BCS Hamiltonian $H_0$ on a square lattice 
is invariant under this transformation. Moreover, if the impurity moments are
aligned along the same direction and there is no potential scattering,
the impurity Hamiltonian $H_{\rm imp}$ will also be invariant under the same transformation.
Potential scattering and randomly oriented impurity moments break particle-hole
symmetry of this nature.

Given that the pairing of electrons occurs in the spin-singlet
state, the superconducting order parameter (amplitude) can be expressed in the form
\begin{equation}
F({\bf R},{\bf r})= \mbox{$1\over 2$} \sum_{\sigma\nu} (i\hat{\tau}_2)_{\sigma\nu}\langle 
\psi_\nu({\bf R}+{\bf r}) \psi_\sigma({\bf R})\rangle. 
\end{equation}
The relation between the order parameter and the gap function is given by the 
equation
\begin{equation}
\Delta({\bf R}) = -U F({\bf R},{\bf r}=0). \label{eq:gap}
\end{equation}
The on-site pairing interaction $U$ is assumed to be instantaneous in time. 
Thus, the energy cutoff in the gap equation is set by the bandwidth. 
In our numerical approach, the gap equation is solved
self-consistently with a given number of quasiparticle excitations
on finite size lattices with periodic boundary
conditions.
In our numerical examples, the strength of the interaction $U$ is chosen 
so that in the absence of impurities and quasiparticle excitations
the energy gap is $\Delta_0/W = 0.1$.

\section{Mapping to an Antiferromagnet}

The Hamiltonian (1) can be transformed to a model where the on-site interaction
is repulsive. In the case of longer-range interactions, Ising-like terms
are generated. On a bipartite lattice, this is achieved by a particle-hole
transformation on the down spins,
\begin{eqnarray}
\psi_\uparrow({\bf r}) &\rightarrow& \psi_\uparrow({\bf r}), \nonumber \\
\psi_\downarrow({\bf r}) &\rightarrow& (-1)^{\bf r}\psi^\dagger_\downarrow({\bf r}) \nonumber.
\end{eqnarray}
In this transformation, the particle number operator transforms to the 
$z$-component of the spin density operator: 
$n({\bf r}) \rightarrow 2s_z({\bf r}) + 1$, and vice versa.
The Hamiltonian is mapped into
\begin{eqnarray}
H&=&-\mbox{$1\over 4$}W \sum_{\langle {\bf Rr}\rangle \sigma}
\psi_\sigma^\dagger({\bf R}+{\bf r}) \psi_\sigma({\bf R}) + h_z\sum_{\bf r} s_z({\bf r}) \nonumber \\
& & \mbox{ } + 2U\sum_{{\bf r}} s_z({\bf r}) s_z({\bf r}),
\end{eqnarray}
where $h_z = U - 2\mu$ is an effective magnetic field along the $z$-axis.
Thus, for the Hubbard model, the particle-hole transformation 
changes the sign of the on-site interaction $U$.

The superconductor has U(1) symmetry associated with the phase of the
order parameter. Because the real and imaginary parts of the order parameter
are transformed to the $x$ and $y$ components of the spin, a gauge
transformation corresponds a rotation of the spin in the $xy$ plane.

The particle-hole transformation establishes one-to-one correspondence
between the ground states of the attractive and repulsive Hubbard
models. For example, consider the attractive Hubbard model away from 
half filling so that the average electron density $\langle n \rangle < 1$
and the average spin density $\langle s_z\rangle = 0$. The particle-hole transformation 
maps it into the half-filled, repulsive Hubbard model with the effective 
magnetic field $h_z$. Its ground state has a transverse antiferromagnetic 
order because in this way the system can lower its energy by generating 
a small ferromagnetic component parallel to the $z$-axis. Therefore, in the 
ground state, $\langle n \rangle = 1$ and $\langle s_z\rangle < 0$. 
Reversing the transformation, the transverse antiferromagnetic order
parameter is mapped to a superconducting order parameter in the attractive
Hubbard model.

Next, consider the attractive Hubbard model away from 
half filling but now in the magnetic field so that $\langle n \rangle < 1$ 
and $\langle s_z\rangle < 0$. This model is mapped by the particle-hole
transformation to the repulsive Hubbard model away from half filling.
It has a ground state which is described by antiphase domain walls 
between antiferromagnetically ordered spins \cite{schulz}. By virtue 
of the particle-hole transformation, it is clear then that the attractive 
Hubbard model with spin-polarized quasiparticles has a superconducting 
ground state where the superconducting domains with the opposite signs of
the order parameter are separated by antiphase domain walls into which 
the excess spin is localized.

\section{The Continuum Model}

Although we are mostly interested in quasi-two-dimensional superconductors,
it is useful to consider one-dimensional systems where many ideas
can be examined analytically. Indeed, for quasi-one-dimensional systems, 
a fruitful connection between the BCS and SSH Hamiltonians can be made. 
The latter one describes, for example, conducting polymers where 
the order parameter $\Delta(x)$ represents the lattice distortion \cite{ssh}. 
In our case, such systems can be organic superconductors or wires whose 
thickness is smaller than the coherence length $\xi_0$. In the weak-coupling 
limit, additional progress is achieved by considering a continuum field theory,
which can be derived because the coherence length is much longer than the
lattice spacing, $\xi_0 \gg a$.

Since we are interested in low-energy and long-wavelength phenomena,
the electronic degrees of freedom can be expressed by slowly varying 
fields $\psi_{\pm\sigma}(x)$ describing the left (+) and right ($-$)
moving electrons,
\begin{equation}
\psi_\sigma(x)/\sqrt{a}= \psi_{+\sigma}(x)e^{ik_Fx} + 
                        \psi_{-\sigma}(x)e^{-ik_Fx},
\end{equation}
where $k_F$ is the Fermi wave vector. Defining the four-component
spinor as $\Psi(x) = [\Phi_\uparrow(x)\ \Phi^\ast_\downarrow(x)]^T$,
where $\Phi_\sigma(x) = [\psi_{+\sigma}(x)\ \psi^\dagger_{-\bar{\sigma}}(x)]^T$,
the BCS Hamiltonian (2) becomes
\begin{equation}
H_{\rm 0}= \int \! dx\, \Psi^\dagger(x)[ v_F\hat{p} - \Delta(x)\hat{\tau}_1]
                      \Psi(x).
\end{equation}
The momentum operator is $\hat{p}=-i\hbar\hat{\tau}_3 \partial_x$, where 
$v_F = (2ta/\hbar)\sin k_Fa$ is the Fermi velocity. Similarly, the gap equation 
can be rewritten in the form \cite{tau}
\begin{equation}
\Delta(x)= -\mbox{$1\over 2$}a U \langle \Psi^\dagger(x)\hat{\tau}_1\Psi(x)\rangle.
\end{equation}
These equations are formally equivalent to those of the TLM model \cite{tlm},
which is the continuum limit of the SSH model. For example, at zero temperature,
the superconducting energy gap is $\Delta_0 = 2W e^{-1/\lambda}$, where the dimensionless
interaction is $\lambda = N_F|U|$; $N_F$ is the density of states at the Fermi
energy in the normal state.  Similarly, the coherence length is
$\xi_0 = \hbar v_F/\Delta_0$.

It is now straightforward to determine the nonequilibrium properties
of the quasi-one-dimensional $s$-wave superconductor. In particular,
it is obvious that injecting spin-polarized electrons into the system,
they form solitons. They are topological excitations of the system,
acting as domain walls between two ground states that differ by the sign
of the order parameter $\Delta$. The energy of the soliton is
$E_{dw}= 2\Delta_0/\pi$ and the order parameter is $\Delta(x)= \Delta_0
\tanh[(x-x_0)/\xi_0]$, where $x_0$ is the location of the center of the
soliton. At low densities of injected electrons, single quasiparticle
bags may appear. They are counterparts of polarons; thus, also their
spatial form as well as their energy, $E_{qp}= \sqrt{2}E_{dw}$,
is known exactly. While in inhomogeneous superconductors individual
quasiparticles may diffuse until they become trapped into defects, 
they are not generic solutions, because at finite concentration of 
quasiparticle excitations they ``phase separate'' forming domain walls.

\section{Antiphase Domain Walls}

While at zero temperature quasiparticle bags are not generic excitations 
of the superconductor,
they may appear as long-lived states because of defects. This may happen
if they are injected into the system at low rate so that they can migrate
without scattering from other quasiparticle excitations long distances 
before they are trapped to defects. Note that, in addition to magnetic 
impurities, a local order-parameter suppression caused by nonmagnetic 
impurities leads to bound states in the superconducting energy gap, 
albeit their binding energies are necessarily small \cite{us2}.
Figure~2 illustrates a situation
which is obtained when the quasiparticle concentration is small and there
are magnetic impurities in the system. For simplicity, the magnetic impurities
are assumed to be ferromagnetically ordered producing a maximal trapping
potential. The bound quasiparticle states yield two peaks in the density of states,
\begin{equation}
{\cal N}(\omega)= -{1\over 2\pi} \sum_{{\bf r}\sigma}
       {\rm Im}\, G_{\sigma\sigma}({\bf r},{\bf r};\omega + i0^+),
\end{equation}
in the energy gap. The oscillations in ${\cal N}(\omega)$
for $|\omega|> \Delta_0$ are due to the finite size effects.

\begin{floating}[t]
\narrowtext
\begin{figure}
\setlength{\unitlength}{1truecm}
\begin{picture}(8.5,12.5)
\put(0.5,1.9){\epsfbox{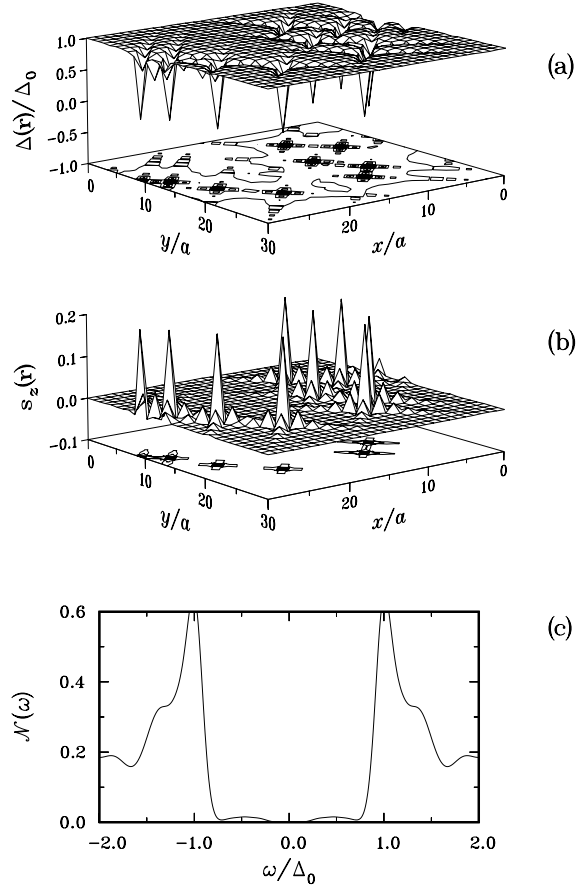}}
\end{picture}
\caption[]{\small
(a) The energy gap, (b) the spin density, and (c) the density of states
of a localized solution of spin-polarized quasiparticles injected 
into an $s$-wave superconductor with magnetic impurities. This configuration 
of well separated quasiparticle bags is obtained self-consistently
on a square lattice with the lattice spacing $a$, $\Delta_0/W= 0.1$, $\pi N_Fw=0.3$,
and $\mu=0$. The concentration of quasiparticles and magnetic impurities
equals 1\%.
}
\end{figure}
\vskip -.5truecm\end{floating}

With an increasing quasiparticle concentration, individual quasiparticle 
excitations become unstable towards a spontaneous formation of antiphase 
domain walls. This tendency is depicted in Fig.~3, where the number of 
quasiparticles is not large enough to form a domain wall that would extend 
all the way through the system. Instead, a closed domain-wall loop is formed.
Because the order parameter changes sign across the domain wall, there 
are midgap states. The finite length of the domain wall leads to the level 
repulsion yielding the density of states that has a minimum at zero energy.
As the system is half filled and either there are no impurities or their 
moments are parallel to each other, the effective Hamiltonian $H_{\rm eff}$ is 
invariant under the particle-hole transformation, Eq.~(\ref{eq:trans}). 
Consequently, the density of states, depicted in Figs.~2 and 3, is symmetric 
relative the zero energy.

\begin{floating}[t]
\narrowtext
\begin{figure}
\setlength{\unitlength}{1truecm}
\begin{picture}(8.5,12.5) 
\put(0.5,1.9){\epsfbox{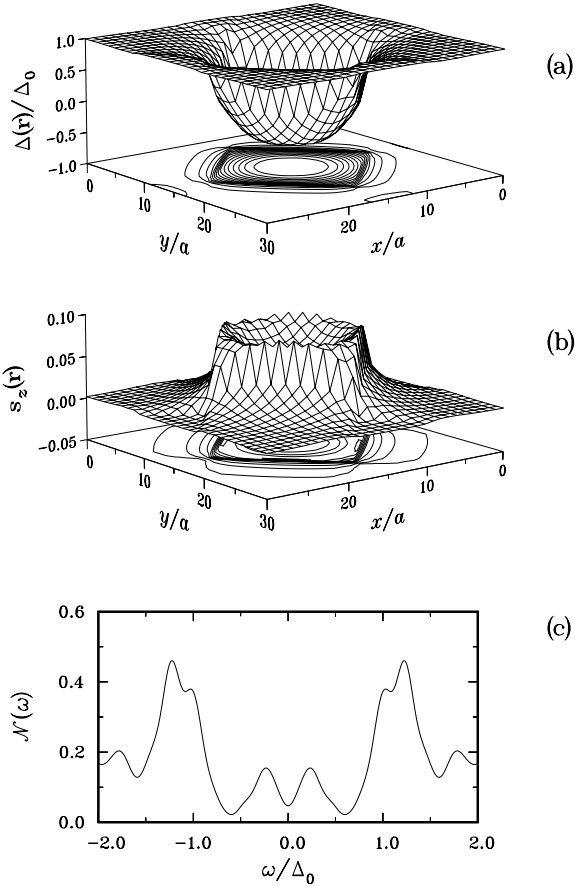}}
\end{picture}
\caption[]{\small
(a) The energy gap, (b) the spin density, and (c) the density of states
when a finite number (40) of spin-polarized quasiparticles is injected 
into an $s$-wave superconductor. This configuration is obtained self-consistently
on a square lattice with the lattice spacing $a$, $\Delta_0/W= 0.1$, and $\mu=0$.
No impurities are present.
}
\end{figure}
\vskip -.5truecm\end{floating}

For a finite concentration of quasiparticles, it becomes energetically 
favorable to form domain walls with infinite length; see Fig.~4. 
This allows all the quasiparticles to occupy the midgap states. 
Domain walls may become pinned to defects either because 
the defects have a magnetic moment or because the defects suppress 
the order parameter locally, and this local suppression then pins
a domain wall. In the case of extended defects, domain walls may
find it preferable to wind through these defects.

\begin{floating}[t]
\narrowtext
\begin{figure}
\setlength{\unitlength}{1truecm}
\begin{picture}(8.5,12.5)
\put(0.5,1.9){\epsfbox{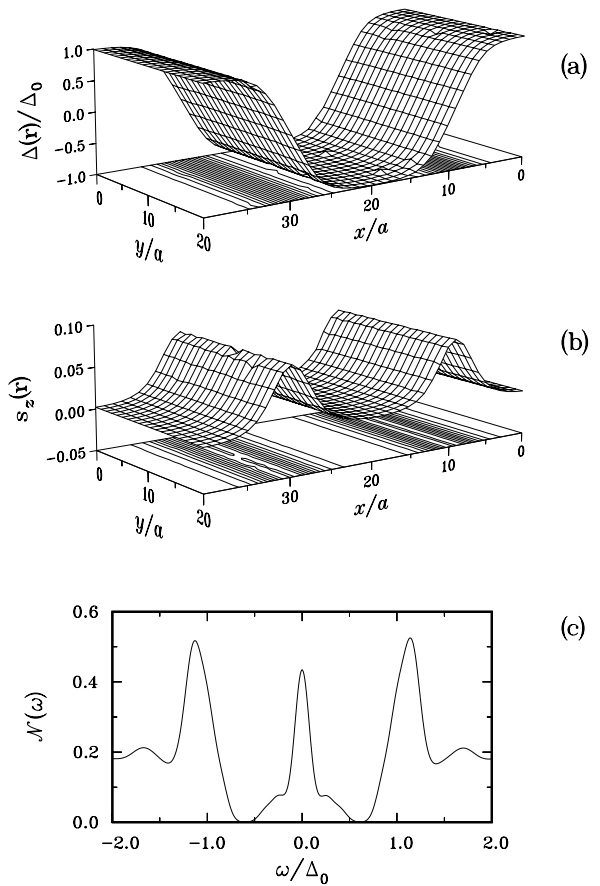}}
\end{picture}
\caption[]{\small
(a) The energy gap, (b) the spin density, and (c) the density of states
in an $s$-wave superconductor with 5\% spin-polarized quasiparticles. 
The antiphase-domain-wall configuration is obtained self-consistently 
on a square lattice with the lattice spacing $a$, $\Delta_0/W= 0.1$, 
$\mu=0$,  $\pi N_FV=0.3$, and $n_{\rm imp}=2\%$.
}
\end{figure}
\vskip -.5truecm\end{floating}

All the quasiparticle and domain-wall textures are charge neutral 
when the system has particle-hole symmetry at the Fermi energy. 
In this regard, quasiparticle bags can be described as spinons \cite{KR}, 
because they carry spin but no charge. On a square lattice with 
the nearest-neighbor hopping, this happens exactly at half filling 
$(\mu=0$). However, if particle-hole symmetry at the Fermi energy 
is broken, self-consistently determined quasiparticle configurations 
usually acquire charge, because they are a linear combination of 
plane-wave states with an energy spread $\Delta\epsilon 
\sim \hbar v_F/\xi_0$ about the Fermi energy. Similarly, away from half 
filling, the domain walls become charged, although their total 
charge per unit length can be quite small. This feature is naturally 
understood by considering the repulsive Hubbard model with a finite 
effective magnetic field which induces a small longitudinal ferromagnetic 
component. In the superconductor, this component is equivalent to 
a non-zero charge density. In contrast, bare quasiparticle excitations 
at the Fermi surface (${\bf k}= {\bf k}_F$) behave as spinons 
irrespective of the energy spectrum in the normal state.

Finally, consider the stability of domain-wall solutions against
a formation of isolated quasiparticle bags. Their energies per particle 
can be computed numerically. In two dimensions, the energy of 
a vertical domain wall per particle is estimated as $E_{dw}
\simeq 0.66\Delta_0$ and the energy of a quasiparticle bag as
$E_{qp} \simeq 0.86\Delta_0$. These estimates are in agreement
with those computed in the antiferromagnetic system for vertical
domain walls \cite{schulz} and spin polarons \cite{su}. Thus, 
approximately at the temperature $T_\ast \sim \Delta_0/5$ 
a considerable fraction of domain walls begins to evaporate 
forming isolated quasiparticle bags. It is interesting to compare 
this temperature with the critical temperature of the superconductor, 
which is $T_c \sim \Delta_0/2$. Thus, there is a sizable temperature 
regime below $T_c$ where most of the excitations appear as isolated
quasiparticle bags. At low enough temperatures, $T\lesssim T_c/3$,
domain-wall textures are thermodynamically favored over non-topological 
quasiparticle configurations.

\section{Optical Absorption}

The optical absorption provides a specific probe to various inhomogeneous
states of nonequilibrium superconductors. 
The optical absorption is the real part of the complex conductivity,
$\sigma'_{ab}(\omega)= {\rm Re}\,\sigma_{ab}(\omega)$ ($a,b=x,y$), where
\begin{equation}
\sigma_{ab}(\omega) = -{1\over i\omega}\big[\Lambda_{ab}({\bf q}=0,\omega+i0^+)
 + {ne^2\over m^\ast}\delta_{ab}\big].
\end{equation}
The ratio between the density of charge carriers and their effective mass
is defined as $n/m^\ast = -a^2\langle H_{\rm kin} \rangle/2$, where
$H_{\rm kin}$ is the kinetic-energy part of the Hamiltonian $H$.
The current-current correlation function is given by the formula
\begin{equation}
\Lambda_{ab}({\bf q},t)= -\langle T j_a({\bf q},t)
j_b(-{\bf q},0)\rangle,
\end{equation}
and its Fourier transform is
\begin{equation}
\Lambda_{ab}({\bf q},\omega) = {1\over N} \int^\infty_0\! dt\, e^{i\omega t}
\Lambda_{ab}({\bf q},t).
\end{equation}
The current operator in Heisenberg picture is defined as 
$j_a({\bf q},t) = e^{iHt}j_a({\bf q}) e^{-iHt}$, where 
\begin{equation}
j_a({\bf q}) = \mbox{$i\over 4$}e a W\sum_{\langle {\bf Rr}\rangle} 
\Psi^\dagger({\bf R}+{\bf r}_a)\hat{\tau}_0 \Psi({\bf R}) e^{-{\bf q}\cdot {\bf r}}.
\end{equation}
It is useful note that optical absorption obeys the sum rule:
\begin{equation}
\int^\infty_0 d\omega \, \sigma'_{aa}(\omega) = {\pi \over 2} {ne^2\over m^\ast}.
\end{equation}
It is a quantity describing any state that is linearly perturbed by 
the electric field. Typically, it is associated with the equilibrium state.
For non-equilibria states, such as the domain walls and quasiparticle bags,
the sum rule must be modified.

\begin{floating}[t] 
\narrowtext
\begin{figure}
\setlength{\unitlength}{1truecm}
\begin{picture}(8.5,5.5)
\put(1.0,-5.0){\epsfbox{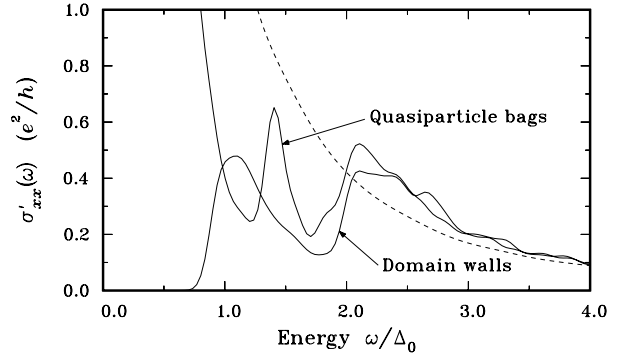}}
\end{picture}
\caption[]{\small
Optical absorption $\sigma'_{xx}(\omega)$ in an $s$-wave superconductor 
in the presence of quasiparticle excitations forming a domain-wall
lattice and localized quasiparticle bags. The concentration of 
quasiparticles in both cases is 1\%. These configurations are 
obtained self-consistently in one dimension for $\Delta_0/W= 0.1$, 
$\mu=0$, $\pi N_FV=0.3$, and $n_{\rm imp}=5\%$. The dashed line 
denotes the optical absorption (Drude-like) obtained in the normal 
state ($\Delta_0= 0$).
}
\end{figure}
\vskip -.5truecm\end{floating}

\begin{floating}[t]
\narrowtext
\begin{figure}
\setlength{\unitlength}{1truecm}
\begin{picture}(8.5,5.5)
\put(1.0,-5.0){\epsfbox{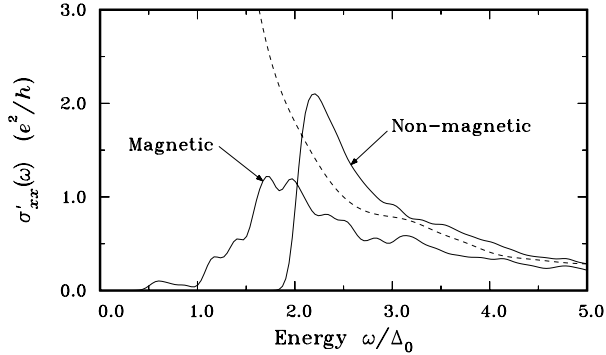}}
\end{picture}
\caption[]{\small
Optical absorption $\sigma'_{xx}(\omega)$ in an $s$-wave superconductor 
with 5\% scalar ($\pi N_FV=0.6$) and magnetic ($\pi N_Fw=0.6$) 
impurities in the absence of quasiparticle excitations. 
The ground-state configuration is obtained self-consistently 
in one dimension for $\Delta_0/W= 0.1$ and $\mu=0$. The dashed line
denotes the optical absorption (Drude-like) obtained in the normal 
state ($\Delta_0= 0$).
}
\end{figure}
\vskip -.5truecm\end{floating}

In the normal state, the optical conductivity has the Drude form
\begin{equation}
\sigma'_{aa}(\omega)= {ne^2\tau\over m^\ast} { 1\over (\tau\omega)^2 + 1},
\label{eq:drude}
\end{equation}
where $\tau^{-1}$ is the scattering rate due to the impurities. In the limit
of dilute concentration of impurities, it can be approximated as
\begin{equation}
\tau^{-1}= {2 n_{\rm imp}\over \pi N_F} \sin^2 \delta,
\end{equation}
where $\delta$ is the phase shift for $s$-wave scattering and $n_{\rm imp}$ 
is the impurity concentration. For point-like impurities, the scattering 
phase shift is obtained from the equation $\cot \delta = c$, where 
$c= (\pi N_F V)^{-1}$, for nonmagnetic impurities, and $c= (\pi N_F w)^{-1}$, 
for magnetic impurities. Below, as a reference, the numerically determined 
optical conductivity in the presence of randomly distributed impurities in the
normal state ($\Delta_0=0$)
is also shown. It is well described \cite{note1} by the Drude form, 
Eq.~(\ref{eq:drude}).

In the superconductor, the zero-temperature optical conductivity typically 
has a threshold of $2\Delta_0$ due to the superconducting energy gap in 
the electronic spectrum at the Fermi energy. Quasiparticle bags and antiphase 
domain walls introduce states within this energy gap that can be used as 
a characteristic signature
of them. Figure 5 illustrates the optical absorption when the
injected quasiparticles form either isolated bag states pinned to
nonmagnetic impurities or domain walls. In the former case, there is a
very large peak in the absorption that comes from the excitation processes
from the localized bag states to states just below the energy gap. 
Because the order-parameter suppression occurs on the length scale
determined by the coherence length, it acts as an attractive potential
with a finite range that can bind states just below the energy gap. 
Thus, in addition to the state occupied by the quasiparticle, the order-parameter
relaxation may admit additional discrete states below the energy gap.
Transitions between these states have a very large oscillator strength.
One can also see a peak at $2\Delta_0$, which is due to the pair-breaking
processes across the energy gap. In the case of domain walls, the optical 
absorption begins at $\Delta_0$ due to the midgap states. One can therefore clearly
distinguish between these two nonequilibrium states based on the optical
absorption.

Impurities yield a quite different absorption spectrum in the absence
of quasiparticle excitations; see Fig.~6. Nonmagnetic impurities produce
a spectrum that has a clear threshold near $2\Delta_0$ and a peak, whereas
magnetic impurities yield a relatively smooth absorption profile which may 
extend deep below $2\Delta_0$, depending on the coupling strength
between electrons and impurity moments. For a high enough concentration of
magnetic impurities, the superconductor becomes gapless and the absorption
will begin at zero energy.

\section{Final Remarks}

Based on both analytical and numerical approaches, we have demonstrated
that $s$-wave superconductors driven away from equilibrium exhibit 
interesting topological textures. They develop as quasiparticles
in nodeless superconductors segregate forming antiphase domain walls 
in the superconducting order parameter and in this manner induce low-energy 
excitations into which quasiparticles relax. Their inhomogeneous
structure has clear experimental implications. For example, a nonuniform 
spin density associated with domain walls should be accessible 
to any probe that is sensitive to a spatially varying magnetization. 
Moreover, optical absorption provides another unambiguous tool for exploring
these textures.

We have assumed that the lifetime $\tau_\ast$ of the quasiparticles in 
the excited state is much longer than the scattering time $\tau_s$ so that
a metastable state is reached. This will require the use of spin-polarized
quasiparticles, which may not always be feasible. A qualitatively similar
situation may be created by maintaining a steady state of unpolarized 
quasiparticles by continuously pumping quasiparticles into excited states.
Even though a genuinely metastable state may not develop because 
$\tau_\ast\sim \tau_s$, the fact that $\tau_\ast$
can be many orders of magnitude longer than the time scale associated with 
the superconducting energy gap, $\hbar/\Delta_0$, suggests that some of 
the features explored here may actually be relevant for such states, too. 
Time resolved techniques are an ideal tool to probe their properties.

While in gapless superconductors, such as in $d$-wave superconductors, it
is no longer clear that quasiparticle excitations will lead to
antiphase domain walls, various external defects that suppress the
order parameter may locally favor a phase shift. Such textures
may appear in magnetic superconductors with static spin-density-wave ordering
where the phases of the magnetic and superconducting order parameters 
intertwine to form a new collective state with midgap quasiparticle 
states.

\section*{Acknowledgments}
 
We would like to thank S.~Kivelson for the interesting suggestion that a ferromagnetic metal
could be used to create and inject spin-polarized quasiparticles into a superconductor.
This work was supported by the NSF under Grant Nos.\ DMR-9527035 and DMR-9629987,
by the U.S.~Department of Energy under the Grant No.\ DE-FG05-94ER45518, 
and by the Many-Body-Theory Program at Los Alamos.


\begin{references}

\bibitem{testardi} L.R.Testardi, Phys. Rev. B{\bf 4}, 2189 (1971).
 
\bibitem{sai} G.A.Sai-Halasz {\it et al.}, Phys. Rev. Lett. {\bf 33}, 215 (1974).

\bibitem{hu} P.Hu, R.C.Dynes, and V.Narayanamurti, Phys. Rev. B{\bf 10}, 
2786 (1974).

\bibitem{iguchi} I.Iguchi, Phys. Rev. B{\bf 16}, 1954 (1977).

\bibitem{fuchs} J.Fuchs {\it et al.}, Phys. Rev. Lett. {\bf 38}, 919 (1977).

\bibitem{dynes}  R.C.Dynes, V.Narayanamurti, and J.P.Garno, Phys. Rev. Lett.
{\bf 39}, 229 (1977).

\bibitem{bob} J.R.Schrieffer and D.M.Ginsberg, Phys. Rev. Lett. {\bf 8}, 207 (1962).

\bibitem{doug} S.B.Kaplan {\it et al.}, Phys. Rev. B{\bf 14}, 4854 (1976).

\bibitem{SK} S.A.Kivelson (private communication).

\bibitem{KR} S.A.Kivelson and D.S.Rokhsar, Phys. Rev. B{\bf 41}, 11693 (1990).

\bibitem{as} A.R.Bishop,~P.S.Lomdahl,~J.R.Schrieffer,~and S.A.Trugman, 
Phys. Rev. Lett. {\bf 61}, 2709 (1988).

\bibitem{coffey} D.Coffey, L.J.Sham, and Y.R.Lin-Liu, Phys. Rev. B{\bf 38}, 5084 (1988).

\bibitem{schulz} H.J.Schulz, J. Phys. (Paris) {\bf 50}, 2833 (1989).

\bibitem{verges} J.A.Verges {\it et al}., Phys. Rev. B{\bf 43}, 6099 (1991).

\bibitem{bcs} J.Bardeen, L.N.Cooper, and J.R.Schrieffer, Phys. Rev. {\bf 108}, 1175 (1957).

\bibitem{ssh} W.-P.Su, J.R.Schrieffer, and A.J.Heeger, Phys. Rev. Lett. 
{\bf 42}, 1698 (1979).

\bibitem{tau} For simplicity, in this section, $\hat{\tau}_1$ denotes 
$\hat{\tau}_1 \otimes \openone$,  which is a block-diagonal matrix equal 
to $\left( { \hat{\tau}_1 \atop 0\phantom{\scriptstyle 0}}{0\phantom{\scriptstyle 0} 
\atop \hat{\tau}_1} \right)$.

\bibitem{tlm} M.Takayama, Y.R.Lin-Liu, and K.Maki, Phys. Rev. B{\bf 21}, 2388,
(1980).

\bibitem{us2} M.I.Salkola, A.V.Balatsky, and J.R.Schrieffer, Phys. Rev. B{\bf 55}, 
12648 (1997).


\bibitem{su} W.P.Su and X.Y.Chen, Phys. Rev. B{\bf 38}, 8879 (1988).

\bibitem{note1} In one dimension, the actual conductivity does deviate from 
the Drude form at low energies because of localization. However, since in our
examples describing $s$-wave superconductors the optical conductivity
vanishes below a small energy scale, this deviation from the Drude behavior
is not shown.

\end{references}
\end{document}